\documentclass{article}

\newcommand{\ls}{\log^{\star}}
\newcommand{\floor}[1]{\scriptstyle \left\lfloor #1 \right\rfloor}

\newcommand{\jth}{\mbox{$j^{\mbox{\scriptsize th}}$ }}
\newcommand{\kth}{\mbox{$k^{\mbox{\scriptsize th}}$ }}
\newcommand{\lth}{\mbox{$l^{\mbox{\scriptsize th}}$ }}

\newcommand{\fo}[1]{f_{\cal S}\left(#1\right)}
\newcommand{\fu}[1]{f'_{\cal S}\left(#1\right)}
\newcommand{\fs}{{\cal F}_{\cal S}}

\renewcommand{\ss}{\mbox{$\cal S$}}
\newcommand{\ee}{\mbox{$\cal E$}}
\newcommand{\oo}{\mbox{$\cal O$}}

\newtheorem{lemma}{Lemma}
\newtheorem{theorem}[lemma]{Theorem}
\newcommand{\proof}[1]{{\bf Proof:}#1 $\diamondsuit$}

\begin{document}

\title{RANDOMIZATION YIELDS SIMPLE $O(n\ls n)$ ALGORITHMS
			FOR DIFFICULT $\Omega(n)$ PROBLEMS\thanks{
				This work has been supported in part by the ESPRIT
				Basic Research Action Nr. 3075 (ALCOM).
				Already published as references
				\cite{prisme-1412a,prisme-1412i,prisme-1412t}.
				}}

\author{Olivier Devillers\\
\small INRIA, 2004 Route des Lucioles, B.P.109\\
06902 Sophia Antipolis cedex (France)}
\date{1991}

\maketitle

\begin{abstract}
We use here the results on the influence graph\cite{idag} to
adapt them for  particular cases where additional information is available.
In some cases, it is possible to improve the
expected randomized complexity of algorithms
from $\scriptstyle O(n\log n)$ to $\scriptstyle O(n\ls n)$.
 
This technique applies in the following applications~:
triangulation of a simple polygon,
skeleton of a simple polygon,
Delaunay triangulation of points knowing the EMST
(euclidean minimum spanning tree).
\end{abstract}

\section{Introduction}
\indent
The classical approach of computational geometry is the search for algorithms
having the best possible worst case complexity. Unfortunately, for difficult
problems, the 
algorithms become fairly complicated and the use of sophisticated data
structures yields unpractical algorithms. Furthermore, the 
authors generally study
only the order of magnitude of the complexity but too complicated algorithms
give implicitely high constants.
For example, although the $\Theta(n)$ time triangulation of a simple polygon by
Bernard Chazelle\cite{triangule_lin} is a beautiful theoretical result,
it does not yield a practical algorithm for
real data on a real computer.

An attractive alternative is to use simpler algorithms
whose complexities are not worst case optimal but only randomized,
i.e.  when averaging over all the possible executions of the algorithm.
In particular, randomized incremental algorithms suppose only that
all the $n!$ possible orders to introduce the $n$ data are
evenly probable. 
It is important to notice that
no hypothesis is made on the data themselves,
so this approach is different from a classical probabilistic point
of view where, for example, the data are supposed
to verify a Poisson's distribution.

The two major techniques for incremental randomized constructions use
the {\em conflict} and {\em influence graphs} respectively.
The conflict graph\cite{claII,mul2,Meh} is a bipartite graph linking
the already constructed results to the data not yet inserted.
The algorithms using such a structure are obviously static,
i.e. the whole set of data must be known in advance
to initialize the conflict graph.

The influence graph is an alternative approach\cite{idag,kdeltree,gks}.
In this structure, all the
intermediate results are linked together to allow the insertion
of further data. The data do not need to be known in advance and can be inserted
on-line, these algorithms are called semi-dynamic.
The analysis is still randomized, but as the data are not known in advance,
they cannot be shuffled and must verify the randomization hypothesis.
More precisely,
the influence graph is a randomized view of an on-line algorithm
and not really a randomized algorithm.
There are also some recent results\cite{remove_Dtree,cms,OSch}
that allow deletions in such structures
and obtain fully dynamic algorithms.

The conflict and influence graphs solve various problems with
optimal expected bounds. For example,
the vertical visibility map of a set of $n$ non intersecting
line segments is computed in $O(n\log n)$ expected time.
In the case where
the segments are connected via their endpoints, Seidel\cite{Seid3}
showed that merging the two kinds of graphs results in a
speed up of the algorithm.
More precisely, the visibility map of a simple $n$-gon is
constructed in $O(n\ls n)$ expected time, using a simple
and practical algorithm.\footnote{
	$\ls n= \mbox{inf} \{k;\log^{(k)}n\leq 1 \}$~;
	for $16 < n \leq 65532 \ , \  \ls n = 4$
	and for $65532 < n \leq 2^{65532} \ , \  \ls n = 5$.
	In other words for any reasonable data set, $\ls n = 4$ and
	for any imaginable computer where addresses are stored with
	less than 65532 bits, $\ls n \leq 5$.
}

This paper proposes simpler proofs for the complexity
of the conflict and influence graphs 
that allow to extend Seidel's technique
to other applications~;
the two main algorithms presented in this paper are the computation of
the skeleton of a simple polygon and the 
Delaunay triangulation of points knowing their euclidean minimum spanning tree.
For the two problems the expected complexity is $O(n \ls n)$,
and the algorithms are simple and easy to code.
The existence of a $o(n\log n)$ deterministic solution is
still open.

\section{Conflict and Influence Graphs}
\subsection{Description of the Problem}
\indent
The problem must be formulated in terms of objects and regions.
The {\em objects} are the input data of the  problem,
they belong to the universe of objects \oo.
For example \oo\ may be the set
of the points, the lines or the hyper-planes of some euclidean space.
The {\em regions} are defined by subsets of \oo\ of less than $b$ objects.
The notion of {\em conflict} is now introduced~: an object and
a region may be, or not, in conflict. If $F$ is a region, the subset
of \oo\ consisting of the
objects in conflict with $F$ is called the {\em influence range}
of $F$.

Now, the aim is to compute for a finite subset \ss\ of \oo, the
regions defined by the objects of \ss\ and without conflict
with objects of \ss~; such a region is called an {\em empty} region of \ss.
The requested result is supposed to be exactly the set of empty regions
or easily deducible from it.

Many geometric problems can be formulated in that way.
The vertical visibility map of line segments is a set of {\em empty} trapezoids.
The Delaunay triangulation of points is a set of triangles with {\em empty}
circumscribing balls.
A visibility graph of a set of line segments is a set of {\em empty} triangles.

\subsection{The Conflict Graph}
\indent
Clarkson and Shor\cite{claII} developed some algorithms based
on a structure called the {\em conflict graph}.
This graph is a bipartite graph between the empty regions
of a subset $\ss'$ of \ss, and the other objects in $\ss \setminus \ss'$.
A region and an object are linked together if they are in conflict.
Thus all the conflict relationships are stored in the conflict graph
and can be used in the algorithm.

The process is initialized with $\ss'=\emptyset$.
There is a unique empty region $\varepsilon$ defined by 0 objects
and each object of \ss\ is in conflict with $\varepsilon$.
At each step, an object $O$
of $\ss \setminus \ss'$ is added to $\ss'$.
All the regions in conflict
with $O$ are known, these regions do not remain empty after the insertion
of $O$ and must be deleted. The new empty regions
defined by $O$ (and other objects of $\ss'$) are created and the
conflicts involving these new regions are computed to replace the conflicts
involving the deleted regions.
When $\ss=\ss'$, the conflict graph is precisely the set of
empty regions of \ss\ which  is exactly the result.

For the randomized analysis,
the points of \ss\ are supposed to be added to $\ss'$ in random order.

\subsection{The Influence Graph}
\indent
The conflict graph gives
immediately the regions in conflict with the new object,
but its design itself requires to know all the objects at
the beginning of the execution.
The algorithms using such a structure are intrinsically static.

The influence graph\cite{idag} is a location structure
for the determination of conflicts.
The nodes of this graph are the regions having been empty at one step
of the incremental construction. This graph is rooted, directed and  acyclic~;
the leaves of this graph are the currently empty regions.
The influence graph satisfies the following property~:
the influence range of a region is included in the union
of the influence ranges of its parents.

The influence graph is initialized with a single node~: the root,
associated to the region $\varepsilon$ whose influence range is the whole
universe of objects \oo.
When a new object $O$ is inserted, the above property
allows to traverse all the regions of the graph in conflict with $O$~;
all the empty regions in conflict are reported.
These regions do not remain empty (they contain $O$) but they still
are nodes of the influence graph.
Then, as for the conflict graph, the new empty regions are computed,
and are linked to the already existing regions
 in order to ensure the further determination of conflicts~;
 they are linked in such a way that
the influence range of a new region is included in the union
of the influence ranges of its parents.

\subsection{Update Conditions}
\indent
For the sake of simplicity, we will make some hypotheses.
These hypotheses are
not really necessary and are
relaxed in\cite{idag}~; but
they are fulfilled by a large class of geometric problems
and allow to express the results in a simple way.
\begin{itemize}
\item Given a region $F$ and an object $O$,
	the test to decide whether or not $O$ is in conflict with $F$
	can be  performed in constant time.
\item If the new object $O$ added to the current set is
	found to be in conflict with $k$ empty regions
	then the computation of the new empty regions requires  $O(k)$ time.
\item In the influence graph, the parents of the new regions
	can be computed in $O(k)$ time and the number of sons of a node is bounded.
\item In the conflict graph, let $O$ be the new object and let $k'$ be the
	number of conflicts between the empty regions in conflict with $O$ and the
	objects not in the current set. Then the computation of conflicts between
	newly created regions and the objects not in the current set
	can be done in time $O(k')$.
\end{itemize}

\section{Analysis\label{Analyse}}
\indent
The classical analysis of these techniques uses random sampling
to bound the number of regions in conflict with at most $k$ objects,
and deduces time and space bounds for the above algorithms.
We propose here a simple analysis where only  bounds for the number of
empty regions and for the number of regions with a single conflict are needed.
In various applications these bounds
can be computed directly, without using
random sampling techniques.

Let us recall here, that all results are randomized, that is
the $n!$ possible orders for the insertion of the $n$ objects in \ss\ are
evenly probable.

We denote~:
\begin{itemize}
\item
	$\omega$ an event, i.e. one of the $n!$ orders.
\item
	$X_{k,l,F}(\omega)$ is 1 if region $F$ is created by the insertion 
	of the \kth object (empty at stage $k$)
	and is in conflict with the \lth object,
	 0 otherwise (always 0 if $k\geq l$).
\item
	$Y_{k,l,F}(\omega)$  is 1 if region $F$ is empty at stage $k$
	and is in conflict with the \lth object, 0 otherwise.
\item
	$X_{k,l}(\omega)=\sum_F X_{k,l,F}(\omega)$
	is the number of conflicts between the regions
	created by the insertion of the \kth object
	and the \lth object in the order $\omega$.
\item
	$Y_{k,l}(\omega)=\sum_F Y_{k,l,F}(\omega)$
	is the number of conflicts between the regions
	empty at stage $k$ and the \lth object.
\item
	For a random sample of \ss\ of size $r$, $\fo{r}$  is the expected
	number of regions defined by the $r$ objects of the sample
	and empty (with respect to the sample)
	and $\fu{r}$ denotes the expected number
	of regions defined by the $r$ objects of the sample
	with exactly one conflict (with an object of the sample).
\end{itemize}

\begin{lemma}
\label{Y}
The expected value of $Y_{k,l}$ (with $k<l$) is $\frac{\fu{k+1}}{k+1}$.
\end{lemma}
\proof{
Let $\omega$ be a given ordering on \ss. Suppose that the \lth\ object $O$
is introduced immediately after the $k$ first elements.
$Y_{k,l}(\omega)$ is the number of regions in conflict with $O$.
By averaging over $\omega$, the $k$ first objects plus $O$ may
be any sample of size $k+1$ with the same probability, 
and $O$ may be any element of the sample with probability
$\frac{1}{k+1}$, which yields the result.
}

\begin{lemma}
\label{XX}
The expected number of regions created by the insertion of the \kth
ob\-ject is less than $\frac{b\fo{k}}{k}$.
\end{lemma}
\proof{
Similar to the preceding one.
}

\begin{lemma}
\label{X}
The expected value of $X_{k,l}$ (with $k<l$) is less than
$\frac{b}{k}\frac{\fu{k+1}}{k+1}$.
\end{lemma}
\proof{
If we know that $Y_{k,l,F}(\omega)=1$ then
$X_{k,l,F}(\omega)=1$ provided that one of the objects describing $F$
is the \kth in the order $\omega$~; this is the case  with probability
less than $\frac{b}{k}$ since the number of objects defining $F$
is less than $b$.

We compute the expected value of  $X_{k,l}$. The sum
is over $\fs$, the set  of regions defined by objects of \ss~:
\begin{eqnarray*}
E(X_{k,l}) 
	   & = & \sum_{F\in\fs} E\left(X_{k,l,F} \right) \\
	   & = & \sum_{F\in\fs} P\left(X_{k,l,F}=1 \right) \\
	   & = & \sum_{F\in\fs} \left[
	  		  P\left(Y_{k,l,F}=1\right) P\left(X_{k,l,F}=1|Y_{k,l,F}=1\right)
							\right. \\
	  & & \hspace*{2cm}
							\left.
			+ P\left(Y_{k,l,F}=0\right) P\left(X_{k,l,F}=1|Y_{k,l,F}=0\right)
							\right] \\
	& < & \sum_{F\in\fs} E\left(Y_{k,l,F} \right) \frac{b}{k} + 0 \\
	& = & E\left(Y_{k,l} \right) \frac{b}{k} \\
	& = & \frac{\fu{k+1}}{k+1} \frac{b}{k}
				\mbox{\hspace*{1cm}using Lemma \ref{Y}}
\end{eqnarray*}
}

\begin{theorem}
\label{general theorem}
The complexity of the operations on
the influence and conflict graphs are the following~:
\begin{enumerate}
	\item \label{size conflict graph}
		The expected size of the conflict graph at stage $k$ is
			$(n-k)\frac{\fu{k+1}}{k+1}$.
	\item \label{update conflict graph}
		The expected number of edges of the conflict graph created at stage $k$
		is less than $\frac{b(n-k)}{k}\frac{\fu{k+1}}{k+1}$
	\item \label{size i-dag}
		The expected size of the influence graph at stage $k$ is less than
			$\sum_{j=0}^k\frac{b\fo{j}}{j}$.
	\item \label{update i-dag}
		The expected cost of inserting the \lth object in the influence graph is less than
			$\sum_{j=0}^{l-1}\frac{b}{j}\frac{\fu{j+1}}{j+1}$.
	\item \label{accelerate i-dag}
		The expected cost of inserting the \lth object in the influence graph
		knowing the conflicts at stage $k$ is less than
			$\sum_{j=k}^{l-1}\frac{b}{j}\frac{\fu{j+1}}{j+1}$.
\end{enumerate}
\end{theorem}
\proof{
\begin{description}
\item[\ref{size conflict graph}]
	The size of the conflict graph is its number of edges. At stage $k$
	the regions present in the conflict graph are exactly the empty
	regions at stage $k$, the number of edges reaching the \jth object
	is $Y_{k,j}$, thus the whole size of the conflict graph is
	$E\left(\sum_{j=k+1}^{n}Y_{k,j}\right)$.
\item[\ref{update conflict graph}]
	An edge of the conflict graph between $F$ and the \jth object
	is created at stage $k$ if $F$ is created at stage $k$ and if $F$
	is in conflict
	with the \jth object. By summing over $j$ we get
	$E\left(\sum_{j=k+1}^{n}X_{k,j}\right)$.
\item[\ref{size i-dag}]
	By the bounded number of sons conditions, the size of the influence
	graph is equal to its number of nodes.
	This number is simply the sum over all
	the regions of the probability for a region to be a node of the graph.
	By Lemma \ref{XX} the expected number of nodes created at stage $j$
	is less than $\frac{b\fo{j}}{j}$.
\item[\ref{update i-dag}]
	During the insertion of the \lth object, the conflicts are located by
	a traversal of the influence graph. A node $F$ is
	visited if it is in conflict
	with the \lth object. By summing over the stage of creation $j$ of $F$
	we get
	$E\left(\sum_{j=1}^{l-1}X_{j,l}\right)$.
	According to update conditions, the number of visited nodes in the
	influence graph is linearly related to the cost of the insertion.
\item[\ref{accelerate i-dag}]
	Same result starting the summation at $j=k$.
}
\end{description}

In the applications described in Section 
5
$f_{\cal S}$ and
$f'_{\cal S}$ are both linear.
In such a case, the complexities get a more explicit expression
stated in the following theorem. Furthermore,
if a direct expression of $f'_{\cal S}$ is not available,
it is possible to show\cite{claII} that 
$\fu{r}=O\left(\fo{\floor{\frac{r}{2}}}\right)$, so it is enough to suppose
that $f_{\cal S}$ is linear.

\begin{theorem}
\label{linear}
If $\fo{r}=O(r)$,
\begin{enumerate}
\item[\ref{size conflict graph}]
	The expected size of the conflict graph at stage $k$ is $O(n-k)$.
\item[\ref{update conflict graph}]
	The expected number of edges of the conflict graph created at stage $k$
	is $O\left(\frac{n-k}{k}\right)$.

	The whole cost of the algorithm is
	$O\left(\sum_{k=1}^{n}\frac{n-k}{k}\right)=O(n\log n)$.
\item[\ref{size i-dag}]
	The expected size of the influence graph at stage $k$ is $O(k)$.
\item[\ref{update i-dag}]
	The expected cost of inserting the
	\lth object in the influence graph is $O(\log l)$.

	The whole cost of the algorithm is
	 $O\left(\sum_{l=1}^{n}\log l\right)=O(n\log n)$.
\item[\ref{accelerate i-dag}]
	The expected cost of inserting the \lth object in the influence graph
	knowing the conflicts at stage $k$ is $O\left(\log \frac{l}{k} \right)$.
\end{enumerate}
\end{theorem}

\proof{
This theorem is simply a corollary of Theorem \ref{general theorem}.
In Point \ref{update conflict graph}, the whole cost can be deduced
because the update conditions ensure that the cost of the algorithm is
related to the total number of structural changes in the conflict graph.
}

\section{Accelerated Algorithms\label{Accelerated algorithms}}
\indent
The principle of accelerated algorithms, introduced by Seidel,\cite{Seid3}
is to exploit Theorem \ref{linear} Point \ref{accelerate i-dag} in order
to achieve a speed up.
The idea is~: {\em if the conflict graph at stage $k$ is
known, the insertion in the influence graph can be done faster}.
At the beginning, the influence graph is constructed in the usual way,
and for some stages $N_i$, the conflict graph at stage $N_i$ is computed
using a direct method exploiting some additional structural
information on the objects. 

To insert the \lth object in the influence graph $(N_i < l \leq  N_{i+1})$,
the conflicts at stage $N_i$ are found using the conflict graph,
and then the conflicts at stage $l-1$ are deduced by traversing
the influence graph.

If $\fo{r}$ is supposed to be $O(r)$,
by choosing
$N_i= \floor{\frac{n}{\log^{(i)}n}}$
(where $\log^{(i)}$ denotes $i$ iterations of  $\log$),
the expected cost of inserting objects in the influence graph between the
key values $N_i$ and $N_{i+1}$ is
\begin{eqnarray*}
\sum_{N_i<j\leq N_{i+1}} O \left( \log \frac{j}{N_i} \right)
& \leq &
\sum_{N_i<j\leq N_{i+1}} O \left( \log \left[ \frac{j}{n} \log^{(i)}n \right] \right) \\
& \leq &
\left( N_{i+1}-N_i\right)  O \left( \log  \log^{(i)}n \right) \\
& \leq &
N_{i+1}  O \left( \log^{(i+1)}n \right) \\
& \leq &
O (n)
\end{eqnarray*}

For an efficient application of this principle, it is necessary
to be able to determine the conflict graph in a direct way
from the whole set of objects, and the set of empty regions
of a sample $r$. We suppose that this can be done in expected time
$O(n)$ (remember that the expected size of this graph is $O(n-r)$).

Thus the expected cost between two key values, for the two steps~:
the influence graph step, and the direct construction of the conflict graph
is $O(n)$. As $N_{(\ls n)-1} \leq n < N_{\ls n}$ the number of relevant
key values is $\ls n$ and the whole expected cost of the algorithm is
$O(n\ls n)$.

\begin{theorem}
\label{Accelerated}
If $\fo{r}=O(r)$, and if the conflict graph between the objects and the
empty regions of a random  sample can be computed in $O(n)$ expected time,
then the accelerated algorithm runs in  $O(n\ls n)$ expected time.
\end{theorem}

\section{Applications\label{Applications}}

\subsection{Triangulating a Simple Polygon}
\indent
The first application is the triangulation of a simple polygon.
This problem can be solved in linear time by a deterministic algorithm
of Chazelle,\cite{triangule_lin} impossible to implement in practice.
Seidel's solution yields a simple randomized algorithm
in $O(n\ls n)$ to compute the vertical visibility map.

This algorithm is not detailed here, the reader can
refer to Seidel's paper.\cite{Seid3}
Seidel's analysis is simpler than that of Section
3
and cannot be generalized directly
because he uses special properties of his application.
More precisely, in Seidel's algorithm
the nodes of the influence graph visited during the
insertion of a new object form a single path.
This fact is used in Seidel's analysis, and yields directly the value of $Y$~:
$\forall\omega, Y_{k,l}(\omega)=1$.

\subsection{Influence and Conflict Graphs for Vorono\"{\i} Diagrams}
\indent
This section presents a randomized algorithm to compute 
the Vorono\"{\i} diagram of a set of points or line segments in the plane
in $O(n\log n)$  expected time. The next sections will be devoted to
accelerated algorithms in $O(n\ls n)$ for special Vorono\"{\i} diagrams.

We consider here the case of the Vorono\"{\i} diagram of
a set of line segments in the plane, for the usual euclidean distance (the dual
of this diagram is called edge Delaunay triangulation~: EDT).  The Vorono\"{\i}
diagram of a set of points is obviously a particular case
and is solved by this algorithm
(a detailed description of this algorithm for points
in any dimensions can be found in a previous paper\cite{deltree}).

We first recall the definition of the Vorono\"{\i} diagram.
\ss\ is a set of objects, here points or line segments in the
euclidean plane \ee. We define the Vorono\"{\i}
cell $V(p)$ of $p\in\ss$ as
$V(p)=\bigcap_{q\in {\cal S}\setminus \{p\} }
				\{m\in\ee;\delta(p,m)\leq\delta(q,m)\}$
where $\delta$ denotes the euclidean distance.

The Vorono\"{\i} diagram $Vor_{\ss}$ is the union of the Vorono\"{\i} cells of
each object of \ss, see Figure \ref{Example} for an example.
	\begin{figure}[htbp]
	\hspace*{2cm}
	\includegraphics{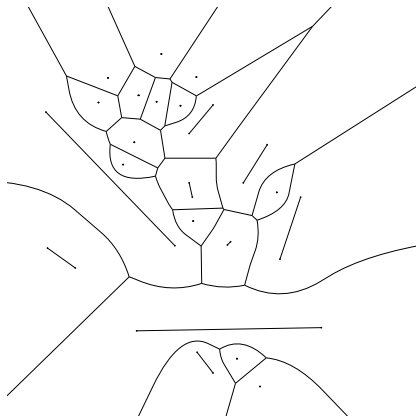}
	\vspace*{6cm}
	\caption{Example of Vorono\"{\i} diagram\label{Example}}
	\end{figure}
These cells  intersect only on their boundaries and form
a partition of the plane.
An important property of the Vorono\"{\i} diagram is that, since
each edge is a portion of a bisecting line,
the maximal empty disk centered on a Vorono\"{\i} vertex touches three objects
and 
the maximal empty disk centered on a Vorono\"{\i} edge touches two objects.

A randomized incremental construction can solve
efficiently the problem of computing $Vor_{\ss}$.
The first point is the definition of objects, regions and conflicts,
such that the Vorono\"{\i} diagram is characterized by the empty regions.
The objects are naturally the line segments (or the points).
As said above, an edge $\Gamma$ of a the Vorono\"{\i} diagram is
a part of a bisecting line of $p$ and $q$, and the endpoints of $\Gamma$
are equidistant to $pqr$ and $pqs$~; such a Vorono\"{\i} edge
(defined by four segments) is called a region and denoted $(pq,r,s)$.
Another segment $m$ is said to be in conflict with $(pq,r,s)$
if $m$ intersects the union of the maximal empty disks centered along $\Gamma$.
In other words, $m$ is in conflict with $(pq,r,s)$ if $\Gamma$
is not an edge of $Vor_{\{p,q,r,s,m\}}$, see Figure \ref{Region}.
	\begin{figure}[htbp]
	\begin{center} 
	
\setlength{\unitlength}{   1.0cm}
\begin{picture}(   5.0,   4.2)(   1.8,   8.7)
\thinlines
\large
\scriptsize
\put(3.25,11.22){$p$}
\put(6.44,10.66){$s$}
\put(4.87,8.95){$q$}
\put(2.12,9.32){$r$}
\put(5.26,12.27){$m$}
\put(4.47,10.18){$\Gamma$}
\put( 1.812, 8.686){\includegraphics{F2.ps}}
\end{picture}
\setlength{\unitlength}{1cm}
	\end{center}
	\caption{$\scriptstyle m$ is in conflict with region $\scriptstyle (pq,r,s)$\label{Region}}
	\end{figure}

In fact it is necessary to be a little more precise in the definition
of regions to hold on some special cases.
Firstly, to describe the unbounded edges of a Vorono\"{\i} diagram,
we just use a new symbol~: $\infty$. The region $(pq,r,\infty)$
corresponds to an unbounded part of the bisecting line of $p$ and $q$,
see Figure \ref{Unbounded1}.
	\begin{figure}[htbp]
	\begin{center} 
	
\setlength{\unitlength}{   1.0cm}
\begin{picture}(   7.2,   5.1)(   2.6,   6.8)
\thinlines
\large
\scriptsize
\put(3.70,10.03){$p$}
\put(5.00,8.22){$q$}
\put(2.80,8.51){$r$}
\put(7.68,11.11){$(pq,r,\infty )$}
\put( 2.552, 6.757){\includegraphics{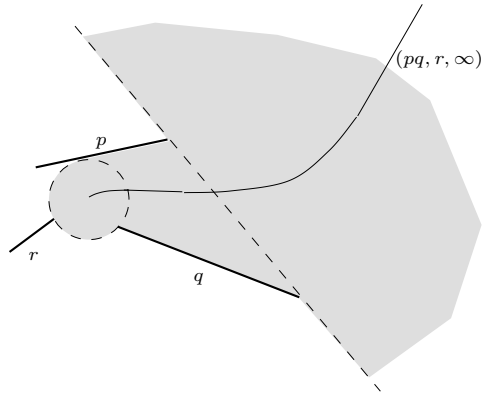}}
\end{picture}
\setlength{\unitlength}{1cm}
	\end{center}
	\caption{An unbounded region $\scriptstyle (pq,r,\infty)$\label{Unbounded1}}
	\end{figure}
Secondly, to ensure the connectivity
of the Vorono\"{\i} diagram, it is necessary
to add some ``virtual'' edges to ``bound'' the unbounded Vorono\"{\i} cells,
see Figures \ref{Unbounded2} and \ref{Ambiguous}.
	\begin{figure}[htbp]
	\begin{center} 
	
\setlength{\unitlength}{   1.0cm}
\begin{picture}(   4.5,   4.4)(   3.9,   7.2)
\thinlines
\large
\scriptsize
\put(6.37,9.66){$p$}
\put(5.68,10.30){$r$}
\put(6.35,10.79){$s$}
\put( 3.915, 7.250){\includegraphics{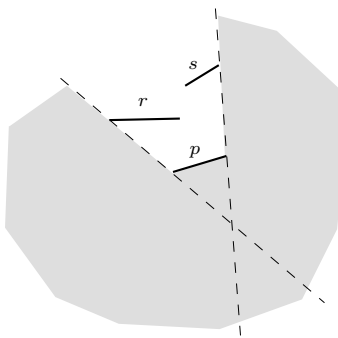}}
\end{picture}
\setlength{\unitlength}{1cm}
	\end{center}
	\caption{Another kind of  unbounded region
					$(p\infty,r,s)$\label{Unbounded2}}
	\end{figure}
Thirdly, in some special cases,
the notation $(pq,r,s)$ may be ambiguous,
but $(pq,r,s)$ can define at most two Vorono\"{\i} edges. In case
of ambiguity, the two regions are distinguished
by the notations $(pq,r,s)^+$ and $(pq,r,s)^-$,
see Figure \ref{Ambiguous}.
	\begin{figure}[htbp]
	\begin{center} 
	
\setlength{\unitlength}{   1.0cm}
\begin{picture}(   7.2,   3.6)(   1.7,   8.7)
\thinlines
\large
\scriptsize
\put(5.55,10.40){$p$}
\put(4.51,11.34){$q$}
\put(4.49,9.57){$r$}
\put(2.35,10.57){$(qr,p,\infty )^-$}
\put(6.52,10.53){$(qr,p,\infty )^+$}
\put(4.73,10.92){$(pq,r,r)$}
\put(4.76,9.89){$(pr,q,q)$}
\put(4.39,12.14){$(q\infty ,r,r)$}
\put(4.28,8.73){$(r\infty ,q,q)$}
\put( 1.653, 8.729){\includegraphics{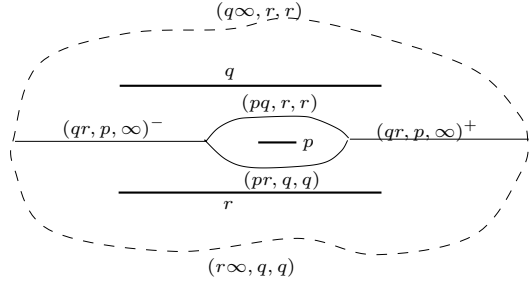}}
\end{picture}
\setlength{\unitlength}{1cm}
	\end{center}
	\caption{Region $(qr,p,\infty)$ can be ambiguous\label{Ambiguous}}
	\end{figure}

It is easy to see that
with these definitions, a region is empty if and only if it corresponds
to an edge of the Vorono\"{\i} diagram.

	\begin{figure}[p]
	\includegraphics{F6.ps}
	\vspace{12cm}
		\begin{center}
			The new segment $\scriptstyle m$ is in bold line.\\
		The dotted edges  correspond to regions in conflict with $\scriptstyle m$.\\
		The dashed edges correspond to new regions created by $\scriptstyle m$.
		\end{center}
	\caption{Insertion of $\scriptstyle m$ in the Vorono\"{\i} diagram\label{EDT1}}
	\end{figure}

The second aspect of the design of a randomized incremental algorithm
is the description of the update procedure for the influence or conflict
graphs.
If a new segment $m$ is added,
the influence graph allows
the  determination of the empty regions in conflict with $m$,
they correspond to disappearing edges of the Vorono\"{\i} diagram.
Consider now $(pq,r,s)$ as a conflicting region. Possibly
one (or even two) portions of the corresponding edge remain in the new diagram,
then the new edge $(pq,r,m)$ for example is made son of $(pq,r,s)$.
So, look at the disappearing part of the Vorono\"{\i} diagram
(see Figure \ref{EDT1}),
it is a tree whose leaves are the vertices of $V(m)$ the new Vorono\"{\i}
cell, they are also the new endpoints of the shortened Vorono\"{\i} edges
described above.
Consider a new Vorono\"{\i} edge on the boundary of $V(m)$
and let $x$ and $y$ be its endpoints. There exists a unique path
of disappearing edges (they form a tree) linking $x$ and $y$.
The new empty region corresponding to the new edge
is made son of all conflicting regions corresponding to edges on this path.
In such a way, an old edge is traversed by two paths
(one for each side of the edge).
A region in conflict with $m$ has at most four sons, two corresponding
to edges on the boundary of $V(m)$ and possibly two shortened edges
(see for example Figure \ref{Ambiguous} and suppose $p$ is inserted last).

For the conflict graph technique, the conflicts with $m$
are directly available and the conflicts of a disappearing region
must be distributed among at most four new regions.

The update conditions are verified, thus, to apply the complexity
results, we just need to know $f_{\cal S}$ and $f'_{\cal S}$.
Here $\fo{n} = O(n)$
because this quantity is related to the size of the order 1 Vorono\"{\i}
diagram\cite{Lee82} and
$\fu{n} = O(n)$
is related to the size of the order 2 Vorono\"{\i} diagram.\cite{Lee82}
The result of Theorem \ref{linear} applies~:
the Vorono\"{\i} diagram (or the edge Delaunay triangulation)
can be computed in $O(n\log n)$ time using the influence graph
(or the conflict graph).

\subsection{Accelerated Delaunay Triangulation Knowing the Euclidean Minimum Spanning Tree}
\indent
It is possible to use the Euclidean Minimum
Spanning Tree of a set of points
(EMST) to speed up the construction of the Delaunay triangulation.
The existence of a deterministic algorithm solving this problem
in $o(n\log n)$ time remains open.
In fact, our technique applies not only for the EMST, but for any
connected spanning subgraph $T$
of the Delaunay triangulation with bounded degree $d$.
All the edges of $T$  are edges of the final Delaunay triangulation.

First, we show that the expected number of intersection points
between $T$ and the Delaunay triangulation of a sample of the points
is $O(dn)$.
Let $vw$ be an edge of $T$ and $ab$ an edge of the Delaunay triangulation
of the sample. There exists an empty region  $abcd$ of the sample.
If $vw$ intersects $ab$, then one of the two points $v$ or $w$ lies
necessarily in the ball circumscribing $abc$ because otherwise,
a circle passing through $v$ and $w$ must contain either $a$ or $b$
and $vw$ cannot be a Delaunay edge in the final triangulation.
So, without loss of generality, suppose that $v$ is in conflict with a region
$abcd$. The number of intersection points between $T$ and $ab$ is bounded
by the number of such points $v$ in conflict with $abcd$ multiplied
by the maximal degree $d$ of a vertex of $T$.
By summing over all regions, the expected number of intersection points is $O(dn)$.

At this time, it is clearly possible to find,
for each vertex of $T$, the Delaunay triangle
in the sample containing the vertex by a simple traversal of $T$
and computing all the intersection points. The other conflicts can be deduced
using the adjacency relations in the Delaunay triangulation of the sample.
Thus Theorem \ref{Accelerated} applies~:
knowing a spanning subgraph of the Delaunay triangulation with maximal degree $d$,
the whole triangulation can be constructed in expected time $O(nd\ls n)$.

The EMST verifies the hypothesis, its edges are in the Delaunay
triangulation\cite{PS} and its maximal degree is less than 6.
(Two edges incident to the same vertex must form an angle
greater than $\frac{\pi}{3}$.)
Thus, knowing the EMST the Delaunay triangulation can be computed in $O(n\ls n)$
expected time.

{
	{\bf Remark~:} if the points are vertices of a convex polygon,
	then this polygon is a correct spanning graph $T$ of degree 2,
	thus the Delaunay triangulation of a convex polygon can be computed in
	$O(n\ls n)$ expected time.
	This problem is solved deterministically
	by Aggarwal et al.\cite{agss}
	using a complicated divide and conquer algorithm
	whose complexity is linear (with a high constant).
	There exists also a still unpublished algorithm by Paul Chew\cite{chew}
	whose randomized expected complexity is linear. The idea is to
	remove the points from the convex hull in the reverse insertion order,
	only maintaining the current convex hull. Thus when inserting the point
	again one conflicting region is known (namely the infinite one) and the
	search for conflicts is avoided.
}

\subsection{Accelerated Skeleton of a Simple Polygon}
\indent
The influence graph can be used to compute the Vorono\"{\i} diagram
of a set of line segments (also called the skeleton).
If these segments form a simple polygon,
or more generally if they are connected then the algorithm can be speed up.

The existence of a deterministic algorithm with complexity $o(n\log n)$
has not already been settled.
Aggarwal et al.\cite{agss} provides an
$O(n)$ deterministic algorithm for a convex polygon,
and Chew's idea\cite{chew} applies also in the special case of a
convex polygon.

Let the line segments $s_0,\ldots,s_{n-1}$ be a simple polygon,
$s_i=p_ip_{i+1}$ ($p_0=p_n$).
For a sample of size $k$, $s_{\sigma{(1)}},\ldots s_{\sigma{(k)}}$,
the Vorono\"{\i} diagram has been already computed.
Then we show how to construct the conflict graph in linear time.

From the line segment $s_{\sigma{(1)}}=p_{\sigma{(1)}}p_{\sigma{(1)}+1}$
the regions defined by $p_{\sigma{(1)}+1}$ are found. Using the adjacency
relations in the Vorono\"{\i} diagram, all the regions in conflict with
$s_{\sigma{(1)}+1}=p_{\sigma{(1)}+1}p_{\sigma{(1)}+2}$
are reported and one region containing point $p_{\sigma{(1)}+2}$ is
kept apart to initialize the search for
the next line segment $s_{\sigma{(1)}+2}$.
By a single walk around the polygon, the whole conflict graph is computed.
The complexity of this algorithm is proportional to the number of
conflicts reported, which is expected to be $O(n)$.

Using Theorem \ref{Accelerated} the skeleton of a simple polygon
(or any connected planar graph) can be computed in
$O(n\ls n)$ expected time.

\section{Conclusion}
\indent
This paper presents various applications of a general scheme of
randomized accelerated algorithms. If a problem can be solved in $O(n\log n)$
time using the usual randomized technique of the conflict graph or the
influence graph, it is often possible to use some additional information
to speed up the algorithm~; by merging both concepts of the
conflict and influence graphs a complexity of $O(n\ls n)$ can be achieved.

This paradigm is applied in Section
5
to two problems having known
deterministic solutions of optimal worst case complexities $\Theta(n)$,
but these algorithms are fairly complicated. These problems are
the triangulation of a simple polygon,\cite{triangule_lin}
and the Delaunay triangulation of a convex polygon.\cite{agss}
In these cases, previous bounds are not improved, but the randomized algorithms
are much simpler.

For the two others applications, no $o(n\log n)$ algorithm was known before.
These problems are the edge Delaunay
triangulation of a simple polygon and
the Delaunay triangulation of a set of points knowing the 
euclidean minimum spanning tree.
Computing the Delaunay triangulation
knowing the EMST in $\Theta(n)$ time will be very interesting
because it will prove the equivalence between the two problems
(the EMST can be deduced from the Delaunay triangulation
in $\Theta(n)$ time).

This technique is powerful and may probably be applied to other problems
whose complexity is $\Omega(n)$ and $O(n\log n)$.

\subsection*{Acknowledgements}
The author would like to thank
 the anonymous referee and other people who communicated to him
	Chew's algorithm,\cite{chew}
 Monique Teillaud and Jean-Daniel Boissonnat
	for a careful reading of the paper
 and Jean-Pierre Merlet for supplying him with
	his interactive drawing preparation system
	{\rm J\kern-.15em\raise.3ex\hbox{\sc p}\kern-.20emdraw }.


\begin{thebibliography}{10}

\bibitem{idag}
J.D. Boissonnat, O. Devillers, R. Schott, M. Teillaud, and M. Yvinec.
  Applications of random sampling to on-line algorithms in
  computational geometry.
  {\it Discrete and Computational Geometry}.
  To be published. Available as Technical Report INRIA 1285. Abstract
  published in IMACS 91 in Dublin.

\bibitem{triangule_lin}
B. Chazelle.
  Triangulating a simple polygon in linear time.
  In {\it IEEE Symposium on Foundations of Computer Science},
  pages~220--230, October 1990.

\bibitem{claII}
K.L. Clarkson and P.W. Shor.
  Applications of random sampling in computational geometry, {II}.
  {\it Discrete and Computational Geometry}, 4(5), 1989.

\bibitem{mul2}
K. Mulmuley.
  On levels in arrangements and {Vorono\"{\i}} diagrams.
  {\it Discrete and Computational Geometry}, 6:307--338, 1991.

\bibitem{Meh}
K. Mehlhorn, S. Meiser, and C. \'O'D\'unlaing.
  On the construction of abstract {Vorono\"{\i}} diagrams.
  {\it Discrete and Computational Geometry}, 6:211--224, 1991.

\bibitem{kdeltree}
J.D. Boissonnat, O. Devillers, and M. Teillaud.
  A semi-dynamic construction of higher order {Vorono\"{\i}} diagrams
  and its randomized analysis.
  {\it Algorithmica}.
  To be published. Available as Technical Report INRIA 1207. Abstract
  published in Second Canadian Conference on Computational Geometry 1990 in
  Ottawa.

\bibitem{gks}
L.J. Guibas, D.E. Knuth, and M. Sharir.
  Randomized incremental construction of {Delaunay} and {Vorono\"{\i}}
  diagrams.
  {\it Algorithmica}.
  To be published. Abstract published in LNCS 443 (ICALP 90).

\bibitem{remove_Dtree}
O. Devillers, S. Meiser, and M. Teillaud.
  Fully dynamic {Delaunay} triangulation in logarithmic expected time
  per operation.
  {\it Computational Geometry Theory and Applications}.
  To be published. Available as Technical Report INRIA 1349. Abstract
  published in LNCS 519 (WADS91).

\bibitem{cms}
K.L. Clarkson, K. Mehlhorn, and R. Seidel.
  Four results on randomized incremental constructions.
  June 1991.
  Manuscript.

\bibitem{OSch}
O. Schwarzkopf.
  Dynamic maintenance of geometric structure made easy.
  In {\it IEEE Symposium on Foundations of Computer Science}, October
  1991.
  Full paper available as Technical Report B 91-05 {Universit\"at}
  Berlin.

\bibitem{Seid3}
R. Seidel.
  A simple and fast randomized algorithm for computing trapezoidal
  decompositions and for triangulating polygons.
  {\it Computational Geometry Theory and Applications}, 1, 1991.

\bibitem{deltree}
J.D. Boissonnat and M. Teillaud.
  On the randomized construction of the {Delaunay} tree.
  {\it Theoretical Computer Science}.
  To be published. Available as Technical Report INRIA 1140.

\bibitem{Lee82}
D.T. Lee.
  On $k$-nearest neighbor {Vorono\"{\i}} diagrams in the plane.
  {\it IEEE Transactions on Computers}, C-31:478--487, 1982.

\bibitem{PS}
F.P. Preparata and M.I. Shamos.
  {\it Computational Geometry~: an Introduction}.
  Sprin\-ger-Ver\-lag, 1985.

\bibitem{agss}
A. Aggarwal, L.J. Guibas, J. Saxe, and P.W. Shor.
  A linear time algorithm for computing the {Vorono\"{\i}} diagram of a
  convex polygon.
  {\it Discrete and Computational Geometry}, 4:591--604, 1989.

\bibitem{chew}
P. Chew.
  A simple randomized linear time algorithm for computing the
  {Vorono\"{\i}} diagram of a convex polygon.
  Unpublished.

\bibitem{prisme-1412t}
O.~Devillers.
\newblock Randomization yields simple {$O(n\log^{\star} n)$} algorithms for
  difficult {$\Omega(n)$} problems.
  \newblock Rapport de recherche 1412, INRIA, 1991.

  \bibitem{prisme-1412i}
  O.~Devillers.
  \newblock Simple randomized {$O(n \log^{*} n)$} algorithms.
  \newblock In {\em Proc. 3rd Canad. Conf. Comput. Geom.}, pages 141--144, 1991.

  \bibitem{prisme-1412a}
  O.~Devillers.
  \newblock Randomization yields simple {$O(n \log^{*} n)$} algorithms for
	difficult {$\Omega(n)$} problems.
	\newblock {\em Internat. J. Comput. Geom. Appl.}, 2(1):621--635, 1992.


\end{thebibliography}

\end{document}